\documentclass[twocolumn,noshowpacs,aps,prl,superscriptaddress]{revtex4}

\usepackage{graphicx}
\usepackage{dcolumn}
\usepackage{amsmath}
\usepackage{epsfig}
\input babarsym

\newcommand{\BABARPubYear}    {03}
\newcommand{\BABARPubNumber}  {038}
\newcommand{\SLACPubNumber} {10295}

\newcommand{\model}{\ensuremath{\mathrm{(model)}}\xspace}

\def\figurebox#1#2#3{%
    \def\arg{#3}%
    \ifx\arg\empty
    {\hfill\vbox{\hsize#2\hrule\hbox to #2{\vrule\hfill\vbox to #1{\hsize#2\vfill}\vrule}\hrule}\hfill}%
    \else
    {\hfill\epsfbox{#3}\hfill}%
    \fi}

\begin{document}

\preprint{\babar-PUB-\BABARPubYear/\BABARPubNumber}

\begin{flushleft}
\babar-PUB-\BABARPubYear/\BABARPubNumber\\
SLAC-PUB-\SLACPubNumber\\
\end{flushleft}

\title{

{\large \bf
Study of high momentum \etapr production in $B\to\etapr X_s$ }


}

\begin{abstract}
We measure the branching fraction for the
charmless semi-inclusive process $B\to \etapr X_s$, where the
\etapr meson has a momentum in the range 2.0 to 2.7
\gevc  in the \FourS center-of-mass frame and $X_s$ represents a system comprising a kaon and zero to 
four pions. We find $\mathcal{B}(B\to \etapr
X_s)=(3.9\pm0.8\stat\pm0.5\syst\pm0.8\model)\times 10^{-4}$. We
also obtain the $X_s$ mass distribution and find that it tends to
favor models predicting high masses.
\end{abstract}


%
\author{B.~Aubert}
\author{R.~Barate}
\author{D.~Boutigny}
\author{F.~Couderc}
\author{J.-M.~Gaillard}
\author{A.~Hicheur}
\author{Y.~Karyotakis}
\author{J.~P.~Lees}
\author{V.~Tisserand}
\author{A.~Zghiche}
\affiliation{Laboratoire de Physique des Particules, F-74941 Annecy-le-Vieux, France }
% ...
\collaboration{The \babar\ Collaboration}
\noaffiliation

\maketitle

\begin{figure}[!htb]
{

\vspace{-0.2cm} }
\begin{center}
\scalebox{0.27}{\includegraphics{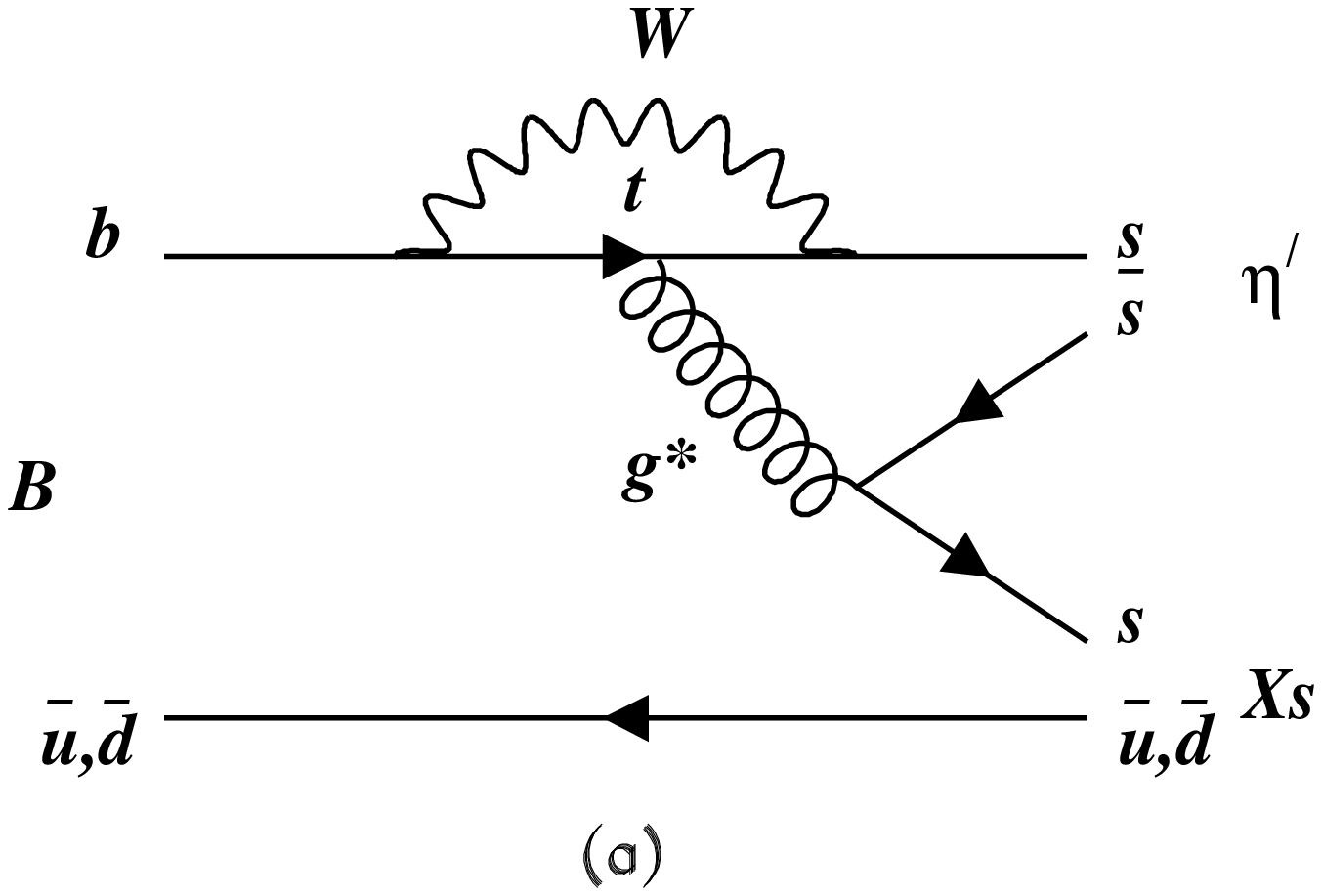}}
\scalebox{0.27}{\includegraphics{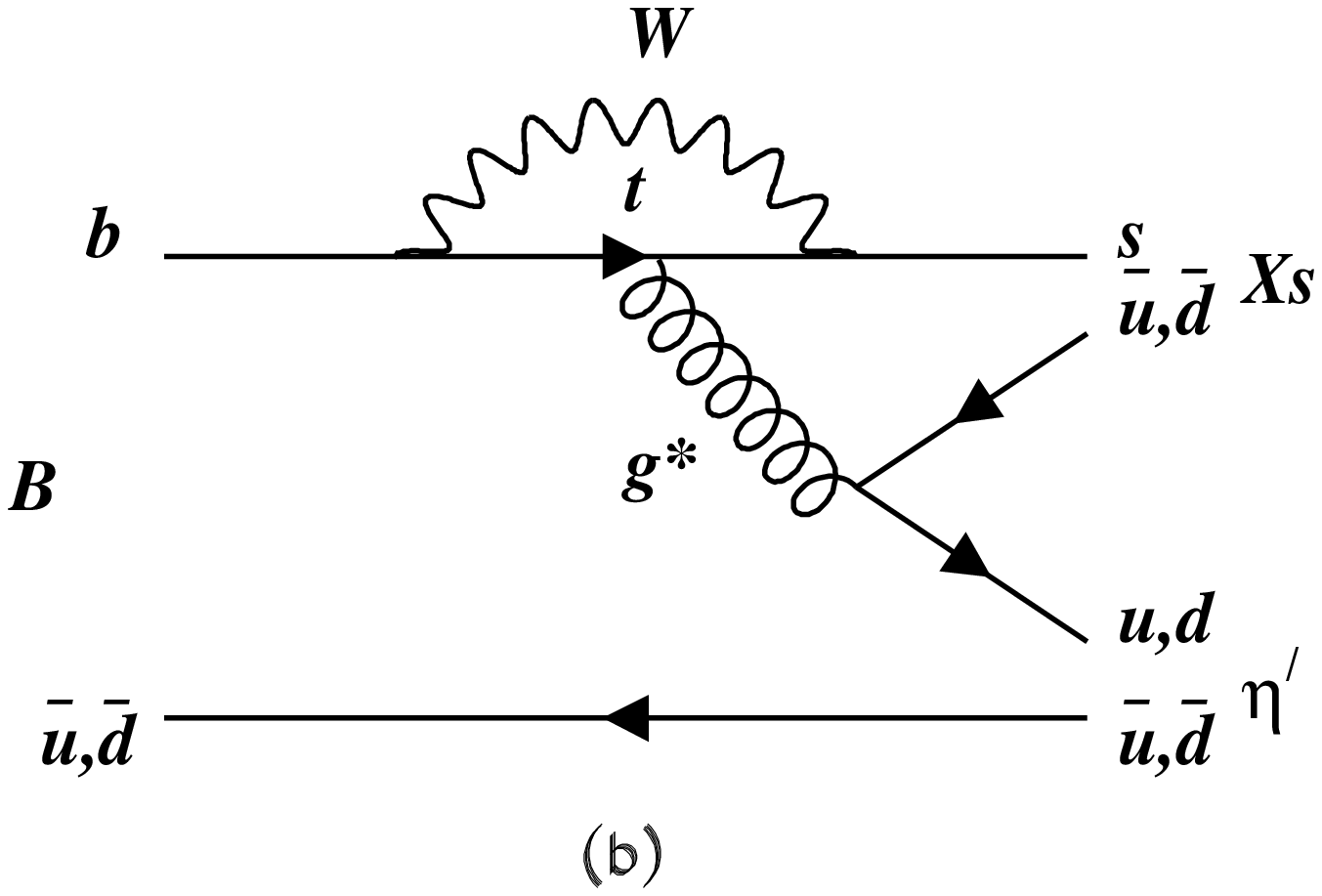}}
\scalebox{0.27}{\includegraphics{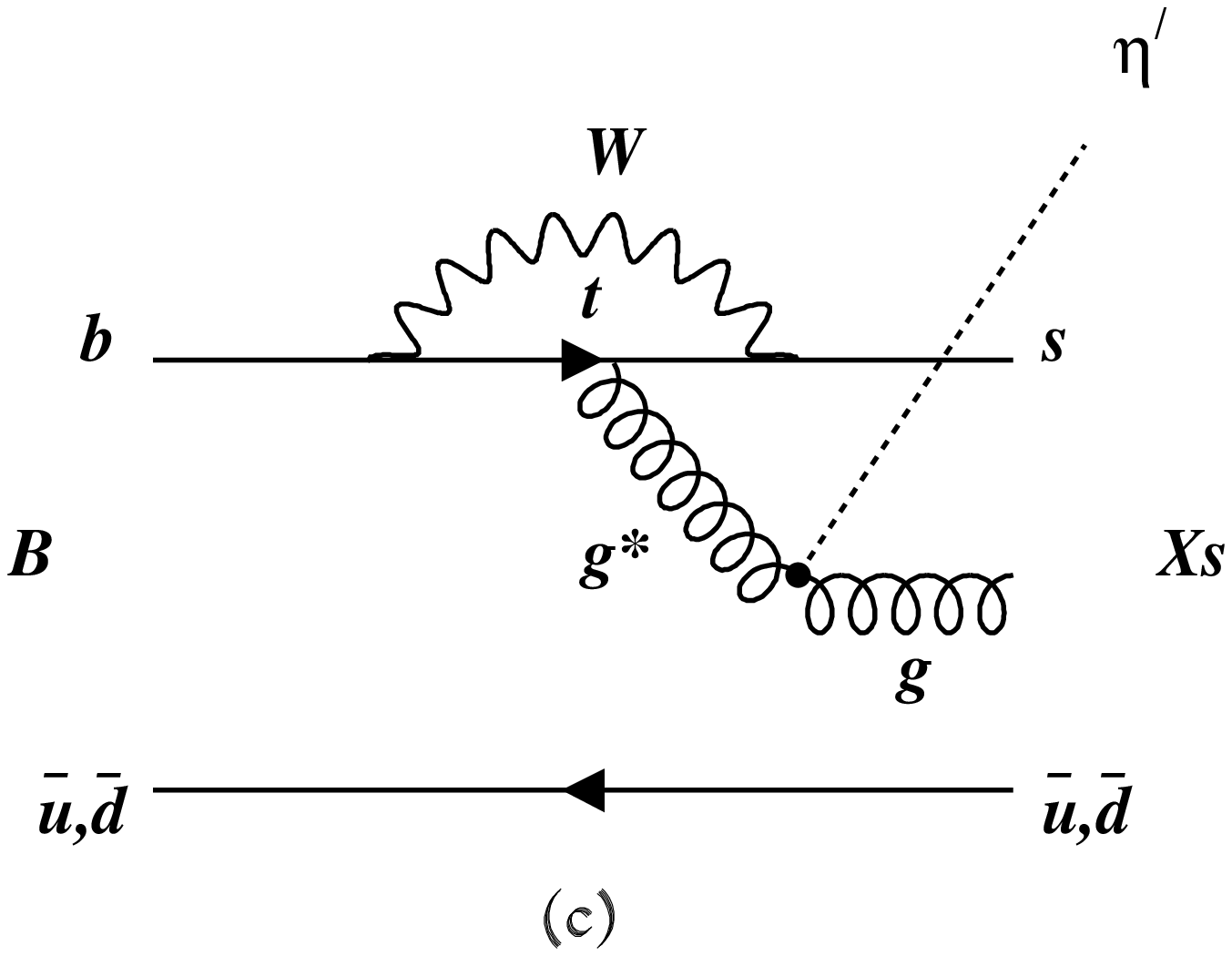}}
\scalebox{0.27}{\includegraphics{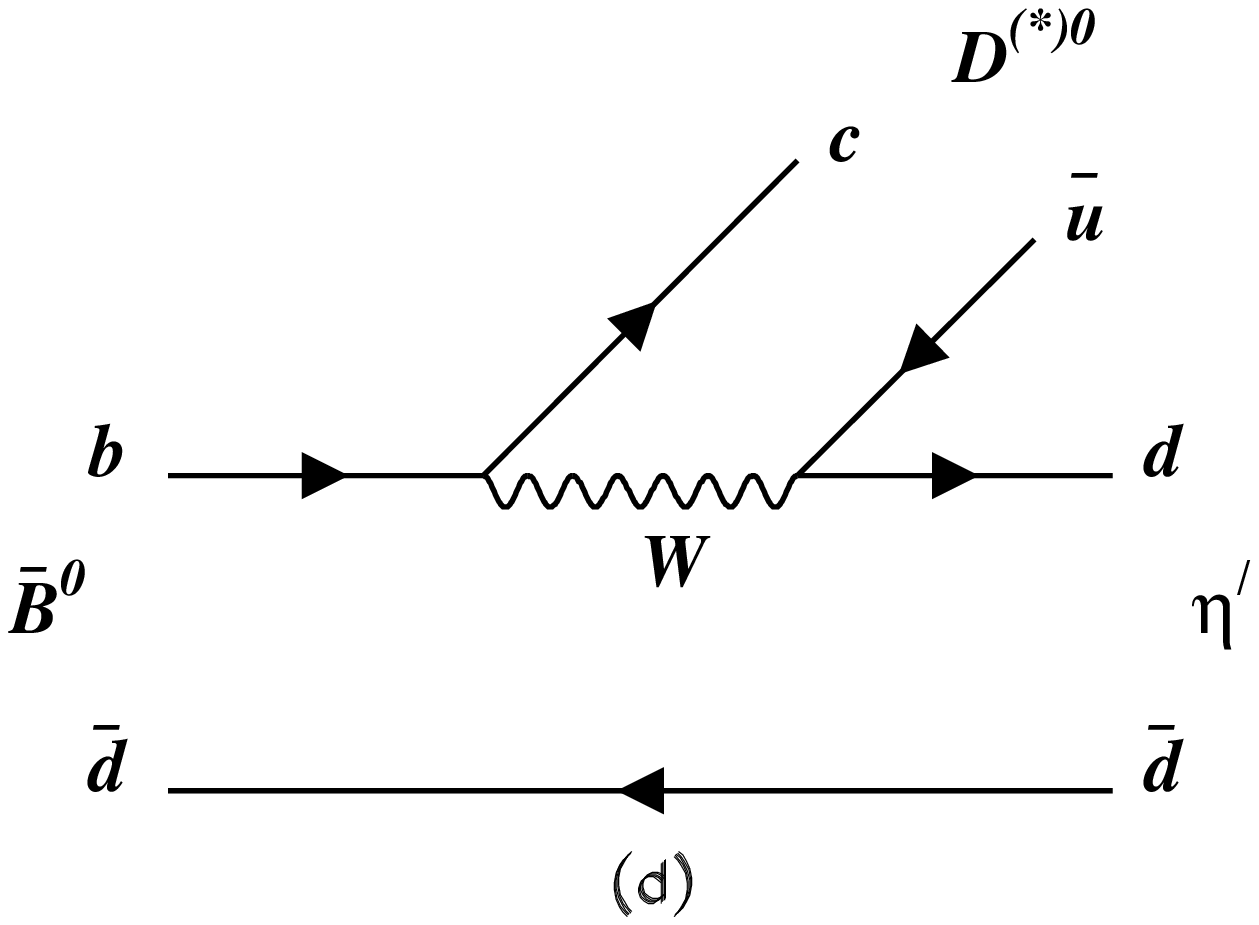}}
\end{center}
\caption{Lowest order diagrams for (a,b,c) $B \to \etapr X_s$ and (d) the color-suppressed background $\Bzb\to\etapr D^{(*)0}$.} \label{fig:diagrams}
\end{figure}

The production of high momentum \etapr mesons in $B$ meson decays is
expected to be dominated by the $B\to\etapr X_s$ process, where $X_s$
is a strange hadronic system, generated by the $b\to sg^*$ transition
as depicted in Fig. \ref{fig:diagrams}(a-c). Figure
\ref{fig:diagrams}(d) shows the color-suppressed modes $\Bzb\to\etapr
D^{(*)0}$, which are 
significant sources of background and which have been measured
for the first time recently \cite{ref:colsup}. Contributions from $b
\rightarrow u$ transitions and other sources of $\etapr$ are expected to be
negligible \cite{ref:BaBarEtaPPi}.

The large $B \to \etapr X_s$ branching fraction measured by the CLEO collaboration \cite{ref:FirstCleo}, 
prompted intense theoretical activity, which focused the special character of the $\etapr$ meson as receiving much of its mass from the QCD anomaly
.  

A later measurement by CLEO confirmed the large \etapr production, measuring $\mathcal{B}(B\to\etapr X_{nc})=(4.6\pm1.1\stat\pm0.4\syst\pm0.5{\rm (bkg)})\times10^{-4}$ \cite{ref:LastCleo}, where $X_{nc}$ denotes a charmless recoiling hadronic system.\\
The rate for $ B \to \etapr X_s$ and especially the fully background-subtracted distribution of the mass of $X_s$ can provide important clues  to the dynamics of weak decays and to the structure of the isosinglet pseudoscalar mesons.\\ 
\indent We present results for the branching fraction $\mathcal{B}(B\to\etapr X_s)$ and the mass spectrum of $X_s$. The signal is analyzed for \etapr momentumbetween  $2.0$  and $2.7~\gevc$ in the CM to suppress background coming from $b\to c\to \etapr$ cascades such as $B\to D_sX$ with $D_s\to\etapr X$, $B\to DX$ with $D\to\etapr X$, $B\to\Lambda_cX$ with $\Lambda_c\to\etapr X$. Our analysis is based on data collected with the \babar\ detector \cite{BABARNIM} at the PEP-II asymmetric $e^+e^-$ collider located at the Stanford Linear Accelerator Center. An integrated luminosity of 81.4 \invfb, corresponding to 88.4 million \BB\ pairs, was recorded at the \FourS resonance (on-resonance) and 9.6 \invfb were recorded 40 $\mev$ below this resonance (off-resonance), for continuum background studies.\\
\indent Two tracking devices are used for the detection of charged particles: a silicon vertex tracker consisting of five layers of double-sided silicon microstrip detectors, and a 40-layer central drift chamber, both operating in the 1.5 T magnetic field of a superconducting solenoid. Photons and electrons are detected by a CsI(Tl) electromagnetic calorimeter. Charged-particle identification is provided by the average energy loss ($dE/dx$) in the tracking devices, and by an internally reflecting ring-imaging Cherenkov detector covering the central region.\\
\indent We select \BB events by requiring at least four charged tracks and a value of the ratio of the second to zeroth Fox-Wolfram moment \cite{ref:fox} less than 0.5.\\
\indent We form a $B$ candidate by combining an
$\etapr\to\eta\pi^+\pi^-$, where the $\eta$ decays into
$\gamma\gamma$, with a $K^+$ or a \KS that is reconstructed in the
$\pi^+\pi^-$ channel, and up to four pions, of which at most one is
a $\pi^0$, leading to 16 possible channels \cite{ref:chargeconj}:
\begin{flushleft}
$B^+\to\etapr K^+(+\pi^0)~~~~~~~~~~~~B^0\to\etapr\KS(+\pi^0)$
$B^+\to\etapr K^+\pi^+\pi^-(+\pi^0)~~~~B^0\to\etapr\KS\pi^+\pi^-(+\pi^0)$\\
$B^+\to\etapr\KS\pi^+(+\pi^0)~~~~~~~~~B^0\to\etapr K^+\pi^-(+\pi^0)$\\
$B^+\to\etapr\KS\pi^+\pi^+\pi^-(+\pi^0)~B^0\to\etapr
K^+\pi^-\pi^+\pi^-(+\pi^0)$

\end{flushleft}
The mass of the $\eta\to\gamma\gamma$, $\KS\to\pi^+\pi^-$ and $\pi^0\to\gamma\gamma$ candidates are required to lie within 3$\sigma$ ($\sigma=16,3$ and $6~\mevcc$ respectively) of their known values and are then kinematically constrained to their nominal masses.\\
\indent To identify the $s$ quark in the $X_s$ system, we require a \KS or a track consistent with a charged kaon. The charged-kaon selection has been optimized to reduce background from $B\to\etapr\pi,~\etapr\rho$, and $\etapr a_1$ decays. For the \KS, we require the angle $\alpha$ between the momentum of the \KS candidate and its flight direction to be less than 0.05 radians, as it peaks at zero for true \KS particles.\\
\indent We require candidates for $B\to\etapr X_s$ to be
consistent with a $B$ decay, based on the beam-energy-substituted
mass,
$\mes=\sqrt{(s/2+\mathbf{p}_0.\mathbf{p}_B)^2/E_0^2-\mathbf{p}_B^2}$
and the energy difference, $\Delta E=E_B^*-\sqrt{s}/2$, where $E$
and $\mathbf{p}$ denote the energy and momentum of the particles,
the subscripts $0$ and $B$ refer to the initial \FourS and the $B$
candidate, respectively, the asterisk denotes the \FourS rest
frame, and $\sqrt{s}$ is the $e^+e^-$ center-of-mass energy.
In addition, the cosine of the angle between the thrust axis of
the $B$ candidate and that of the rest of the event in the
center-of-mass frame ($\cos\theta^*_T$) is used to remove
continuum background, which is peaked near $|\cos\theta^*_T|=1$,
while signal events are uniformly distributed. We require
$\mes>5.265~\gevcc$, $|\Delta E|<0.1~\gev$, and
$|\cos\theta^*_T|<0.8$. For each event, we select the 
candidate
with the smallest $\chi^2$, with  $\chi^2$  defined by
\begin{center}
$\chi^2=(\mes-M_B)^2/\sigma^2(\mes)+(\Delta E)^2/\sigma^2(\Delta
E)$\,,
\end{center}
where $M_B$ is the $B$-meson mass and where
where the resolutions $\sigma(\mes)=3~\mevcc$ and $\sigma(\Delta E)=25~\mev$ are obtained from Monte Carlo simulation. 
The remaining continuum background is subtracted with the use of off-resonance data.\\
\indent The background contribution 
from color-suppressed modes
$\Bzb\to\etapr D^{(*)0}$ is estimated from a Monte Carlo
simulation which uses our measurement of its branching fraction,
$\mathcal{B}(\Bzb\to\etapr
D^{(*)0})=(1.7\pm0.4\stat\pm0.2\syst)\times 10^{-4}$
\cite{ref:colsup}.

\begin{figure}[!btp]
\begin{center}
\scalebox{0.2}{\includegraphics{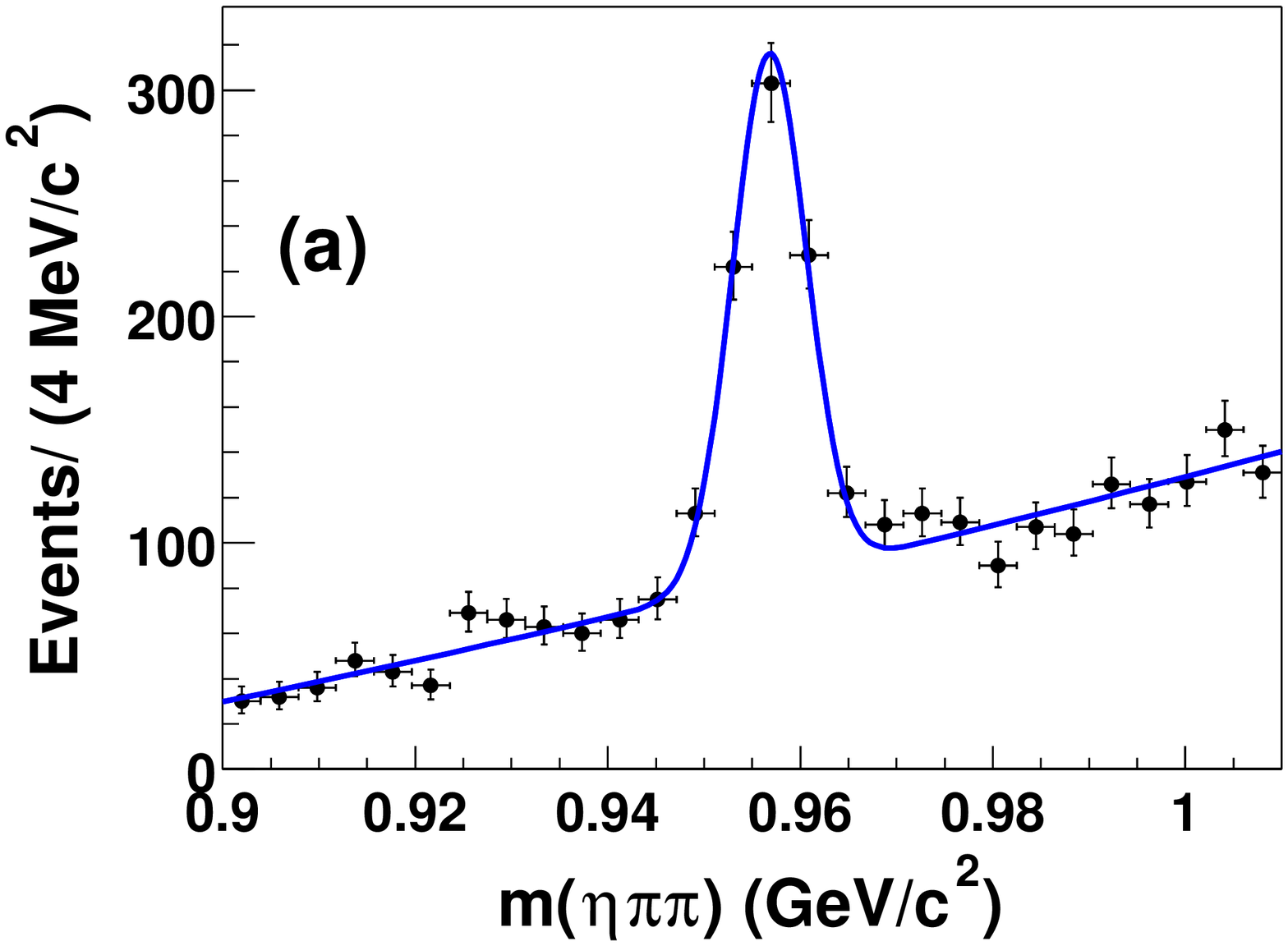}}
\scalebox{0.2}{\includegraphics{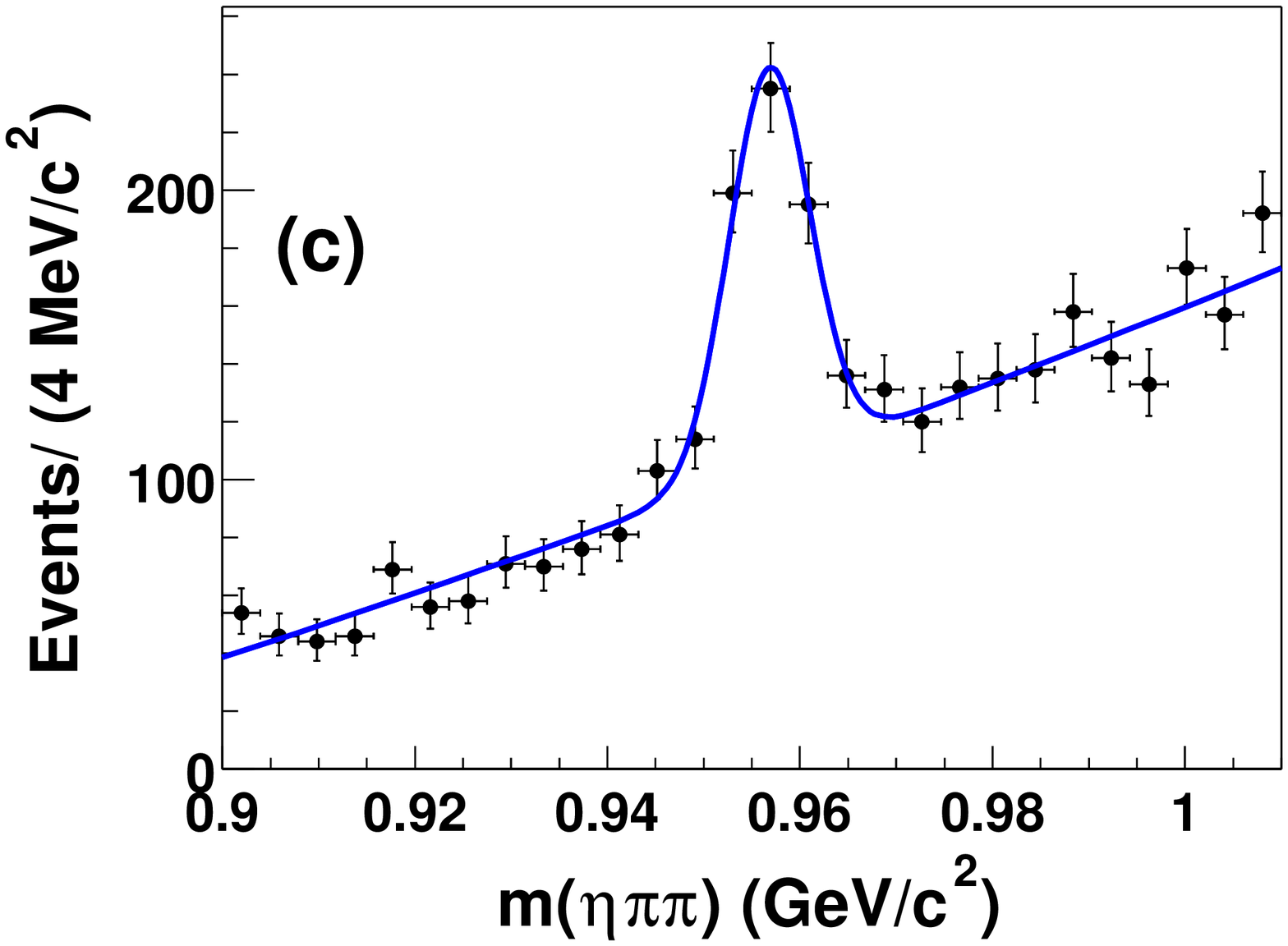}}
\scalebox{0.2}{\includegraphics{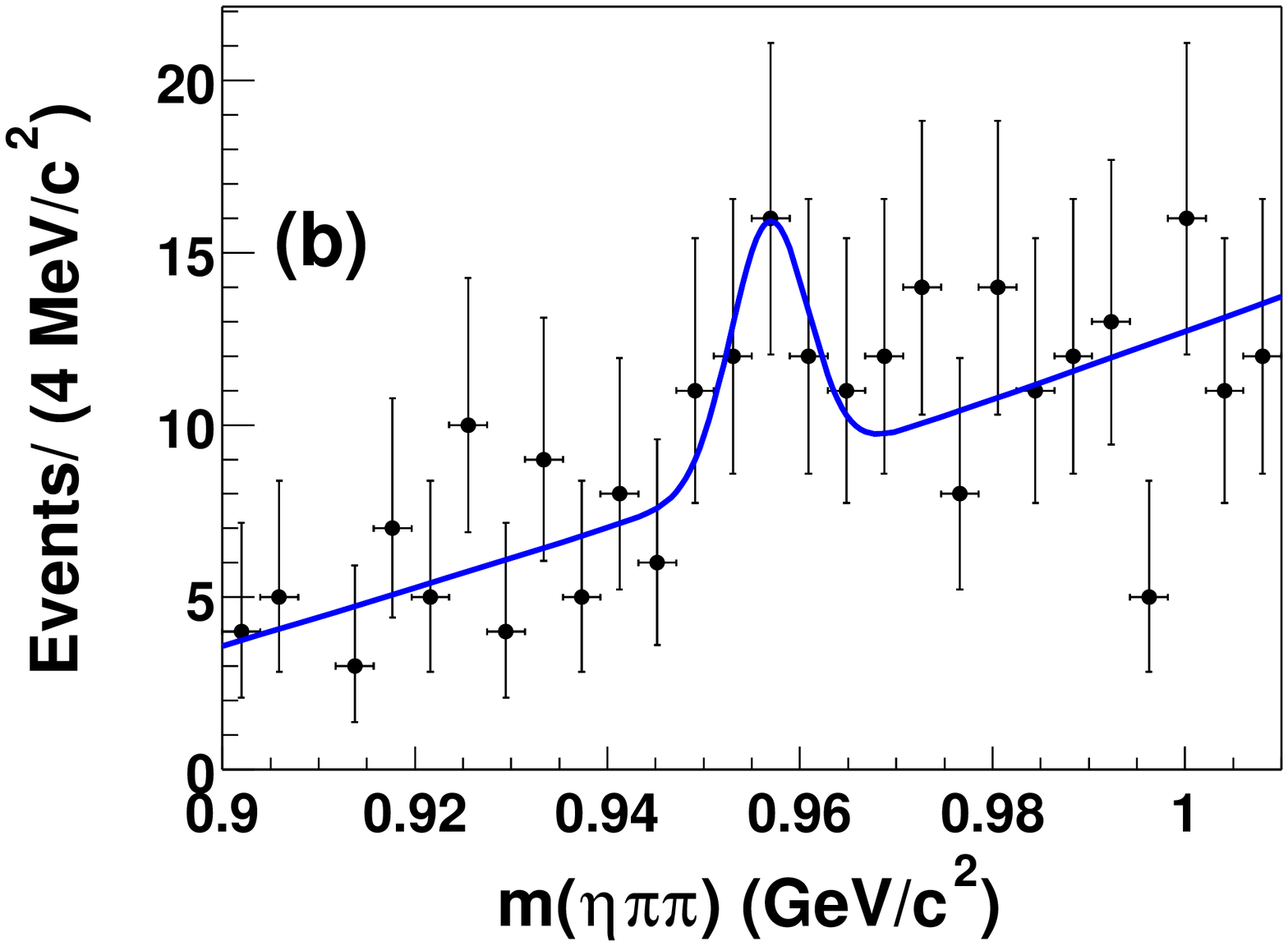}}
\scalebox{0.2}{\includegraphics{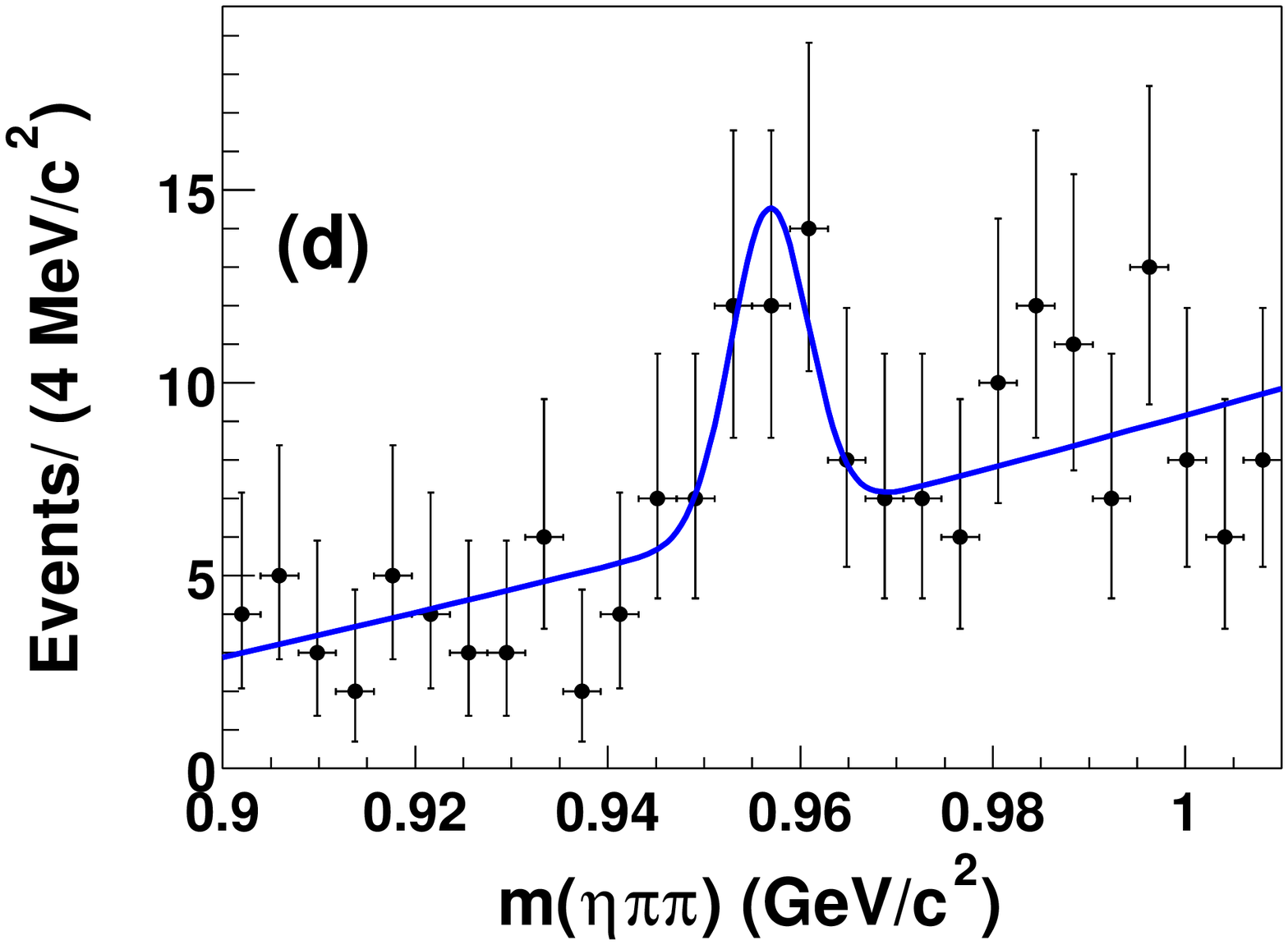}}
\end{center}
\caption{Fits to the $\eta\pi\pi$ invariant mass for on-resonance (top) and off-resonance (bottom) data samples, for the modes (a,b) $K^{\pm}$ and (c,d) \KS.}
\label{fig:etaPrFits}
\end{figure}

To determine efficiencies, we model the signal using a combination of the two-body mode
$B\to\etapr K$ and, for $X_s$ masses above the $K\pi$ threshold, a
non-resonant derived from the 
theoretical predictions \cite{ref:Atw,ref:Hou,ref:Fritzsch}, which are based on the anomalous $\etapr$-gluon-gluon coupling and which favor high-mass $X_s$ systems. The
fraction of the two-body mode is constrained in the simulation model to be between 10\% and
15\% \cite{ref:newBaBarEtaPK,ref:BelleEtaPK}. When not forming a $K$ meson, the $X_s$ fragments into
$s\bar{q}$ and  $s\bar{q}g$ ($q=u,d$). We find that the overall efficiency is
$(6.0\pm0.2)$\% for the $K^{\pm}$ modes and $(4.7\pm0.1)\%$ for the
\KS modes, including the branching fraction
$\mathcal{B}(\KS\to\pi^+\pi^-)$.\\ 
\indent The branching fraction of $B\to\etapr X_s$ is computed through a fit to the number of \etapr signal events, with \etapr momentum between $2.0$ and $2.7~\gevc$, both for on-resonance and off-resonance data. To parameterize the background, we use a Gaussian function for the signal and a second order polynomial. For the fit of the off-resonance data sample, we constrain the mass and width of the \etapr to the values obtained with on-resonance data. Figure \ref{fig:etaPrFits} shows the fits of the $\eta\pi\pi$ invariant mass distributions for the $K^{\pm}$ and \KS modes. The fitted yields are reported in Table \ref{Ta:summary}.\\
\begin{table}[!btp]
\begin{center}
\caption{Results of the fits for $K^{\pm}$ and \KS modes. Yields for on-resonance data ($Y_{\rm ON}$), off-resonance data ($Y_{\rm OFF}$), expectation from color-suppressed background ($Y_{\rm CS}$) and on-resonance data after background subtraction ($Y$) are given. A luminosity scale factor, $f=8.48$ , is applied to the off-resonance yield.}\label{Ta:summary}

\begin{tabular}{|lcc|}
\hline & \hspace{0.5cm}$K^{\pm}$ modes & \hspace{0.5cm}\KS modes \\
\hline $Y_{\rm ON}$ & \hspace{0.5cm}$577.0\pm34.0$ & \hspace{0.5cm}$367.0\pm34.0$\\
$Y_{\rm OFF}$ & \hspace{0.5cm}$18.9\pm8.5$ & \hspace{0.5cm}$21.7\pm8.4$ \\
$Y_{\rm CS}$ & \hspace{0.5cm}$63.6\pm 11.4$ & \hspace{0.5cm}$26.9\pm4.5$ \\
\hline $Y$ & \hspace{0.5cm}$353.1\pm 80.5$ & \hspace{0.5cm}$156.1\pm79.1$ \\
\hline
\end{tabular}
\end{center}
\end{table}
The semi-inclusive branching fraction is computed by performing a weighted average of the results obtained for the $K^{\pm}$ and \KS modes.
The detection efficiencies are corrected to account for the \etapr and $\eta$ branching fractions to the channel we observe. For the \KS modes, we convert the result so it corresponds to $K^0$ and ${\overline K}^0$.
The final state $X_s$ includes both $K^+$- and $K^0$-tagged decays. 
Assuming that their branching fractions are equal, we obtain $\mathcal{B}(B\to\etapr X_s)=(3.9\pm0.8\stat\pm0.5\syst\pm0.8\model)\times 10^{-4}$.
We obtain the systematic error by combining the sources listed in Table \ref{Ta:systExcl}.

\begin{table}[!btp]
\begin{center}
\caption{Contribution of different sources to the systematic error for modes with a $K^{\pm}$ or \KS.}\label{Ta:systExcl}
\begin{tabular}{|lcc|}
\hline
 Source &  $K^{\pm}$ syst (\%)& \KS syst (\%)\\
\hline
Tracking & 3.4 & 3.3\\
$\eta,\pi^0$ detection& 7.0 & 8.2\\
$K/\KS$ ID& 2.5 & 4.3\\
$\mathcal{B}(\etapr\to\eta_{\gamma\gamma}\pi\pi)$ & 3.4 & 3.4\\
$N_{\BB}$ & 1.1 & 1.1\\
MC sample size & 3.0 & 3.0\\
$\etapr D^{(*)0}$ subtraction & 3.0 & 2.9\\
\hline Total & 12.1 & 13.5\\
\hline Model & 20 & 20\\
\hline
\end{tabular}
\end{center}
\end{table}

\begin{figure}[!btp]
{
\hspace{0.2cm}(a)
~~~~~~~~~~~~~~~~~~~~~~~~~~~~~~~~
(b)

\vspace{-1.0cm} }
\begin{center}
\scalebox{0.2}{\includegraphics{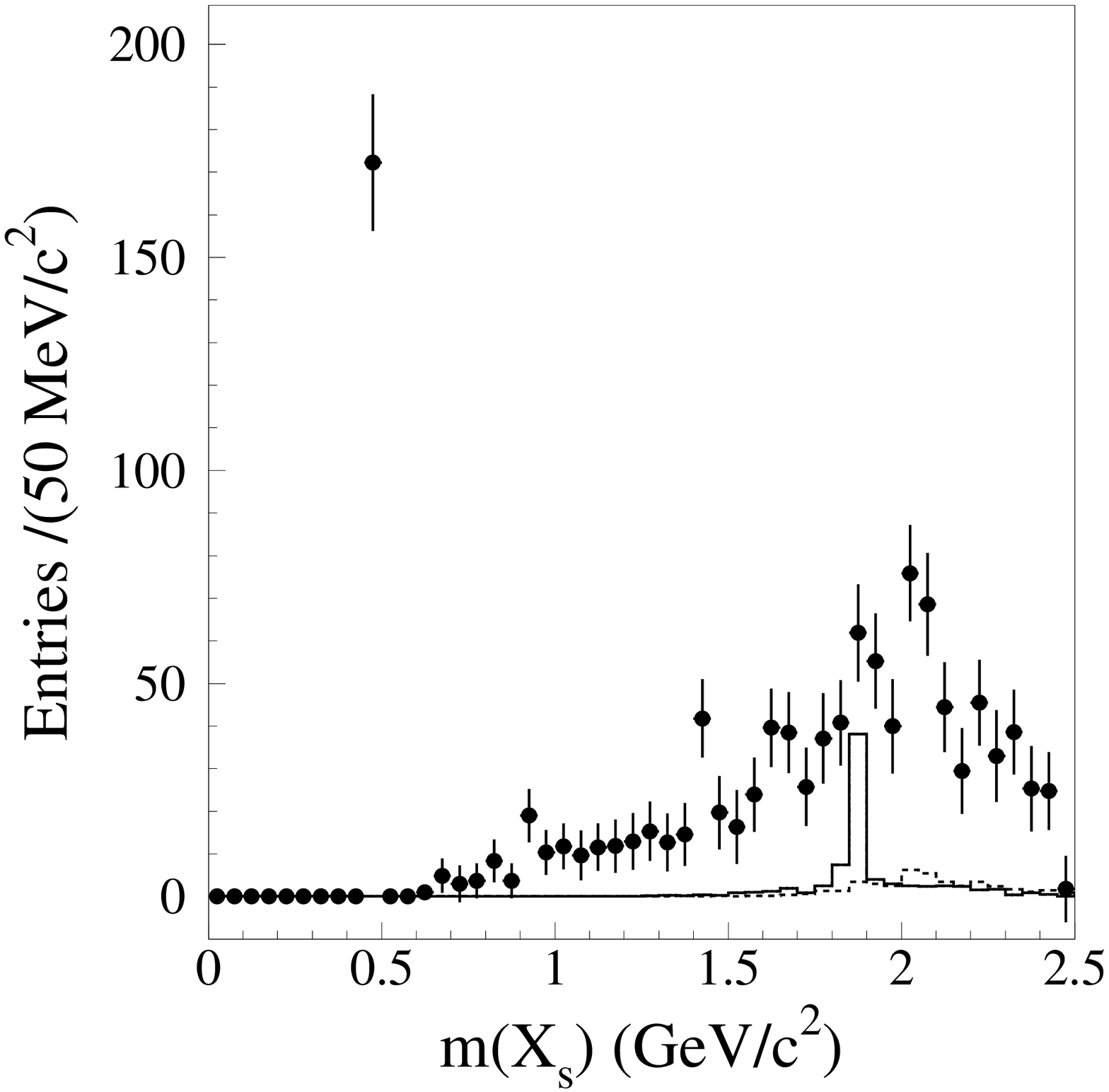}}
\scalebox{0.2}{\includegraphics{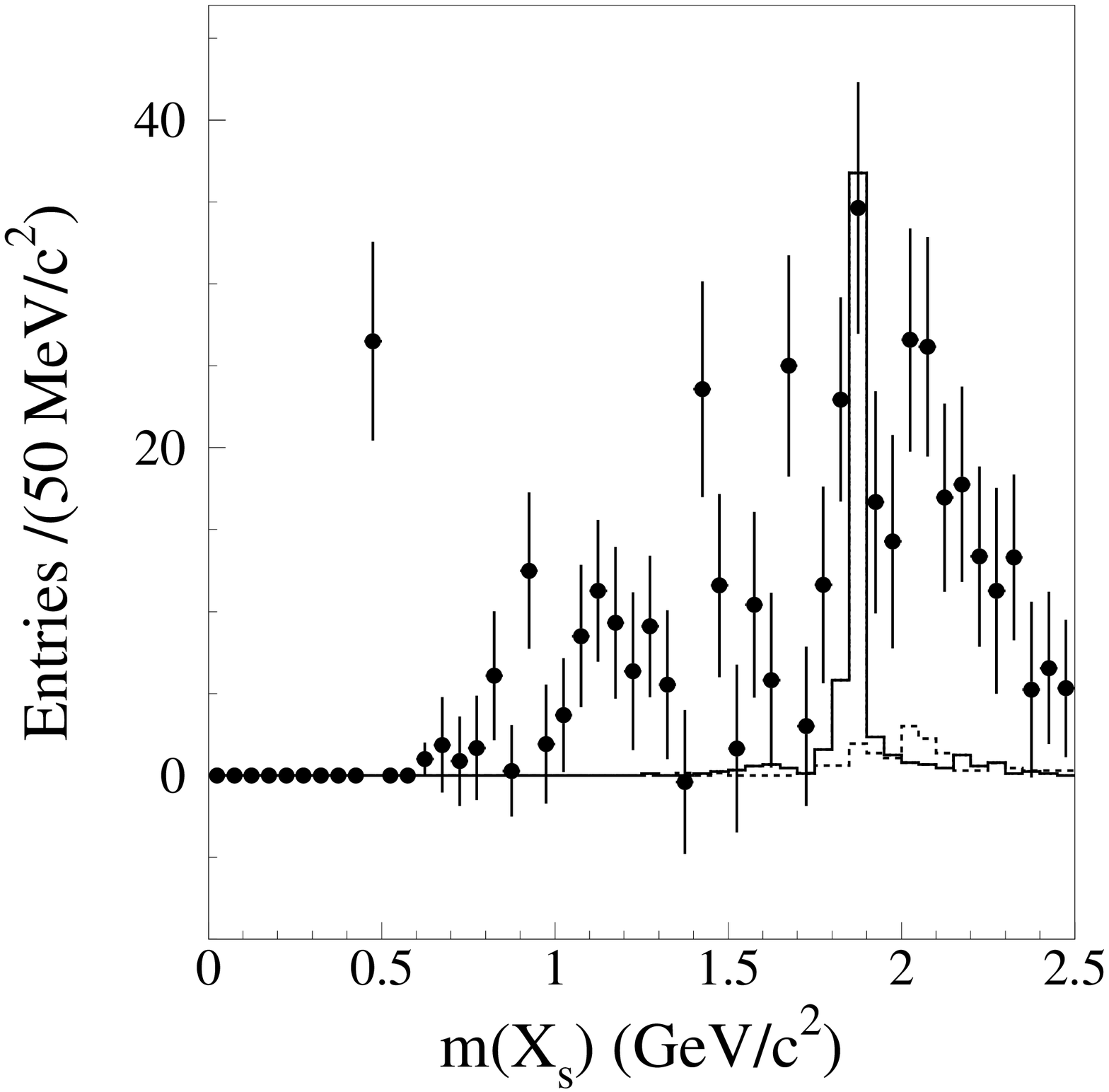}}
\caption{Continuum-subtracted $K\ n\pi$ invariant-mass distributions
for (a) all $B$ modes and (b) $B^0$ modes, including combinatorial
background. Solid and dashed histograms represent expected backgrounds from $\Bzb\to\etapr D^0$ and $\Bzb\to\etapr D^{*0}$,
respectively.}\label{Fi:RawMXs}
\end{center}
\end{figure}

The largest uncertainty arises from our model of the $X_s$ system. 
To estimate that uncertainty, we use an alternative model which consists of a combination of resonant modes:
$\etapr K$, $\etapr K^*(892)$, $\etapr K_1(1270)$, $\etapr
K_1(1400)$, $\etapr K^*(1410)$, $\etapr K_2^*(1430)$, $\etapr
K_3^*(1780)$, and $\etapr K_4^*(2045)$. The variability of the efficiency and our knowledge of the resonant sector lead us to assign a 20\% systematic uncertainty.
 Other systematic uncertainties include track reconstruction efficiency, reconstruction efficiencies of $\pi^0\to\gamma\gamma$, $\eta\to\gamma\gamma$, and $\KS\to\pi^+\pi^-$ candidates, charged-kaon identification efficiency, secondary branching fractions, number of \BB\ events ($N_{\BB}$), the size of our Monte-Carlo sample, and subtraction of the background from $\Bzb\to \etapr D^{(*)0}$.\\
\indent To explore the $X_s$ mass distribution, we select
$B$ candidates for which the mass of the \etapr is within
three standard deviations of the known value, and subtract the
continuum contribution by using on-resonance data in the sideband
$5.200<\mes<5.265~\gevcc$. The continuum background scaling factor ($\mathcal{A}$), from the sideband to signal regions, is computed
from off-resonance data to be $0.591\pm0.118$.
The resulting mass distributions are shown in Fig. \ref{Fi:RawMXs}
for all $B$ modes and separately for the $B^0$ modes. The peak at
$m(X_s)\simeq500 \mevcc$ corresponds to the two body mode $B\to
\etapr K$.

\begin{figure}[!btp]
\begin{center}
\scalebox{0.25}{\includegraphics{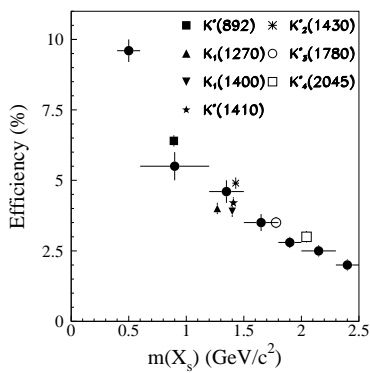}}
\caption{Variation of efficiency with $m(X_s)$. The filled circles indicate the efficiency for non-resonant $X_s$ simulation. The other symbols denote the values for the resonances.}\label{Fi:effmxs}
\end{center}
\end{figure}

To obtain the full 
$X_s$ spectrum, we fit the \etapr mass
distribution in bins of $X_s$ mass. The efficiency, averaged over
the charged and neutral kaons, as a function of $m(X_s)$, is shown
in Fig. \ref{Fi:effmxs}. The correction for the feed-across
between bins is included in the efficiencies.

According to simulations, the $X_s$ system is correctly reconstructed for 85\% (60\%) of the
candidates in the region $m(X_s)<1.5~\gevcc$
($m(X_s)>1.5~\gevcc$). For correctly reconstructed events, the
experimental resolution varies from 5 to 15 $\mevcc$ for low and
high masses, respectively. In the case of misreconstructed events,
the resolution ranges from 100 to 150 $\mevcc$.
Table~\ref{Ta:YieldMxs} shows the fitted yields for the raw
signal, the sideband region, the expected color-suppressed
background, and the yield after full background subtraction, as a
function of $m(X_s)$.

\begin{table}[!btp]
\begin{center}
\caption{Fitted yields for on-resonance data and color-suppressed background for different $m(X_s)$ ranges in \gevcc. The sideband yields ($Y_{SB}$) must be corrected by the sideband to signal region scaling factor (see text) before subtraction.}\label{Ta:YieldMxs}
\begin{tabular}{|lcccc|}
\hline $m(X_s)$ range & $Y_{ON}$ & $Y_{SB}$ & $Y_{CS}$ & $Y$\\
\hline \hspace{0.3cm}$[0.4,0.6]$ & $200\pm15$& $46.1\pm8.8$& --- & $172.8\pm15.9$\\
\hspace{0.3cm}$[0.6,1.2]$ & $120 \pm 14$& $100 \pm 13$& --- & $60.9\pm16.0$\\
\hspace{0.3cm}$[1.2,1.5]$ & $114\pm15$& $112\pm14$& $1.1\pm0.3$& $46.7\pm17.1$\\
\hspace{0.3cm}$[1.5,1.8]$ & $150\pm18$& $163\pm17$& $7.7\pm1.6$& $46.0\pm20.7$\\
\hspace{0.3cm}$[1.8,2.0]$ & $140\pm17$& $93\pm15$& $47.4\pm9.6$& $37.6\pm21.4$\\
\hspace{0.3cm}$[2.0,2.3]$ & $149\pm20$& $142\pm18$& $26.2\pm4.5$& $38.9\pm23.1$\\
\hspace{0.3cm}$[2.3,2.5]$ & $80\pm14$& $70\pm14$& $4.9\pm0.9$& $33.7\pm16.3$\\
\hline
\end{tabular}
\end{center}
\end{table}

\begin{figure}[!btp]
{
\hspace{-1.5cm}(a)
~~~~~~~~~~~~~~~~~~~~~~~~~~~~~~~~~
(b)

\vspace{-1cm} }
  \begin{center}
\scalebox{0.2}{\includegraphics{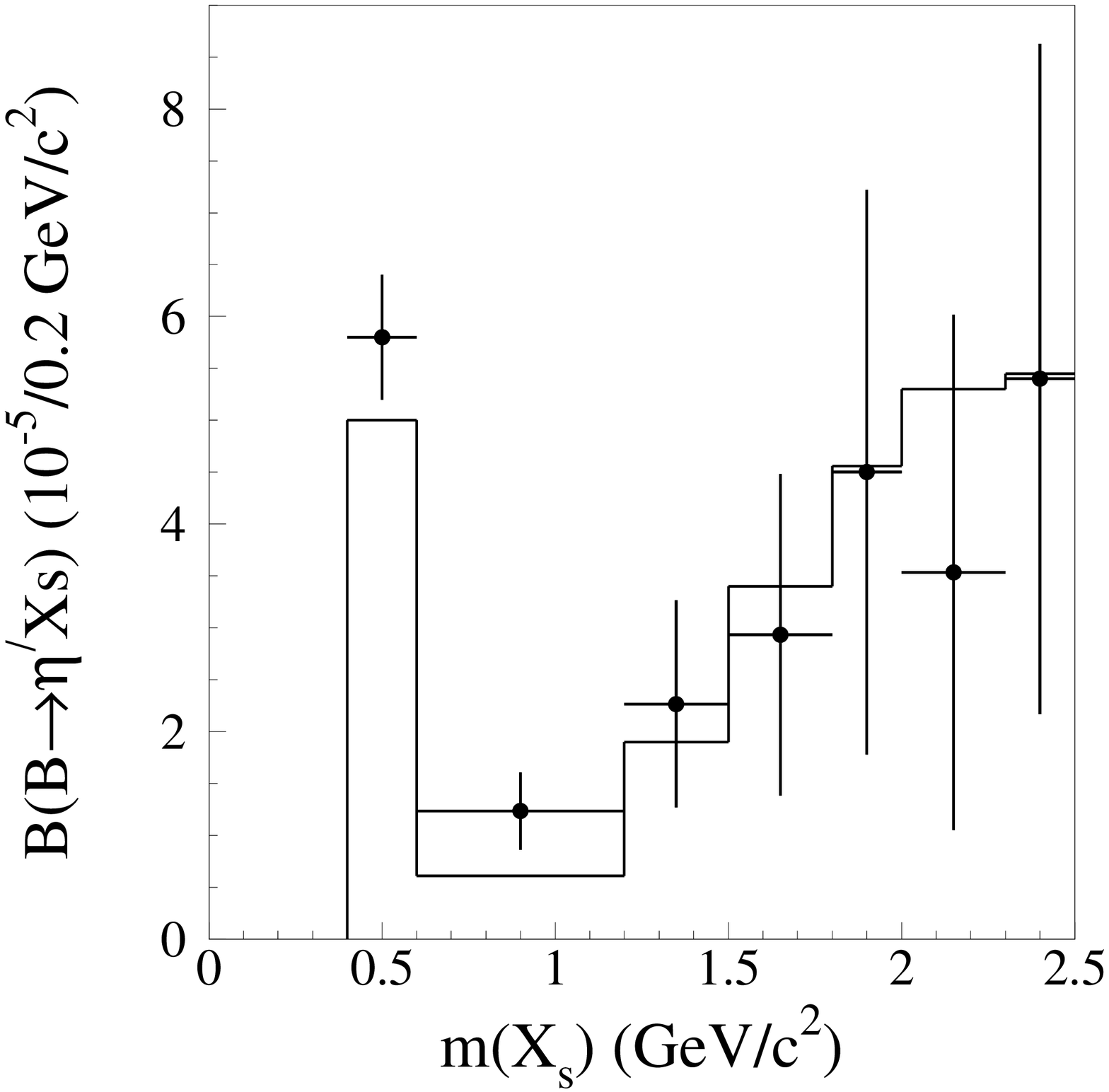}}
\scalebox{0.2}{\includegraphics{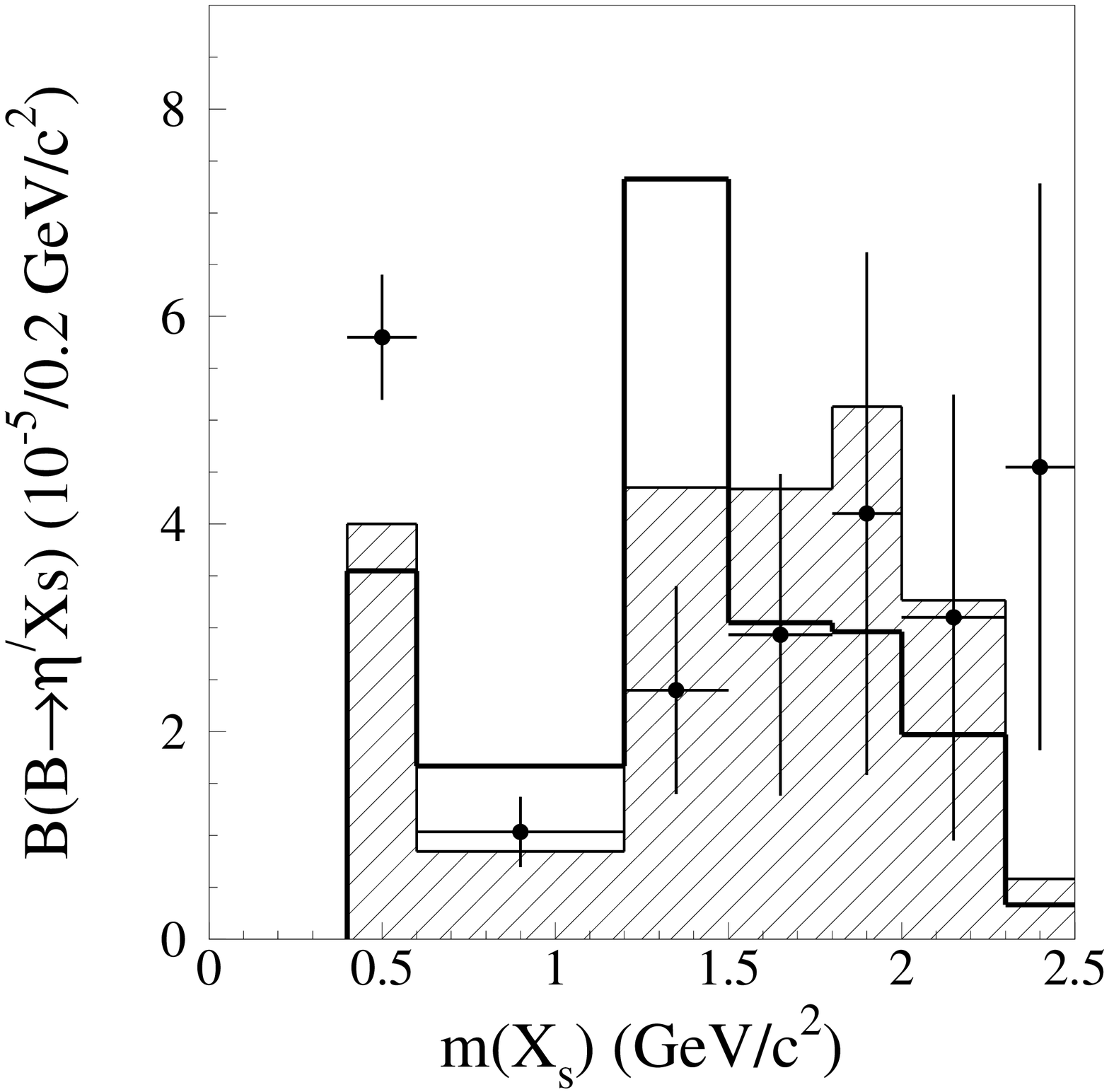}}
\caption{Branching fractions as a function of $m(X_s)$. Both (a) and (b) show the 
same data, though the efficiency used in (a) is derived from the non-resonant model, while that in (b)  the efficiency comes from the model with a combination of resonances. The errors include bin-to-bin systematics; an additional systematic error of $\sim$ 8\% (not shown) is common to all points. (a) The open histogram
represents the expectation from non-resonant $m(X_s)$ simulation.
(b) The open histogram represents the expectation from a
mixture of resonant modes with equal proportions. The hatched histogram results if some heavy resonances are enhanced. The equal mixture provides a good approximation to what is predicted in \cite{ref:Datta}.} \label{Fi:BFVsMxs}
\end{center}
\end{figure}

The branching fraction as a function of $m(X_s)$, obtained from
the fully background-subtracted yield (Table \ref{Ta:YieldMxs}),
is shown in Fig. \ref{Fi:BFVsMxs}.

We compare data and simulation by forming a \chisq difference. The \chisq probability for the nonresonant $X_s$ model (Fig. \ref{Fi:BFVsMxs}(a)) to fit the data is 61\% while it is close to $\sim10^{-7}$ for the equal mixture of resonances (Fig. \ref{Fi:BFVsMxs}(b)). We find improved agreement with the resonant model if the weights of $K_3^*$ and $K_4^*$ are increased by a factor of 1.5, leading to a probability of 2\%.\\
\indent As a consistency check of the method, we measure the two-body decay modes ($X_s=K^{\pm},\KS$), and find $171.0\pm14.0$ and $27.1\pm5.6$ events in on-resonance data for $\etapr K^{\pm}$ and $\etapr\KS$ respectively, and no \etapr signal events for both channels in off-resonance data, leading to the branching fractions $\mathcal{B}(B^{\pm}\to\eta^{\prime}K^{\pm})=(6.9\pm0.6\stat)\times 10^{-5}$ and $\mathcal{B}(B^0\to\eta^{\prime}K^0)=(5.6\pm1.2\stat)\times 10^{-5}$. These values are fully compatible with what has been measured by recent exclusive analyses \cite{ref:newBaBarEtaPK,ref:BelleEtaPK}.\\
\indent In summary, we have measured the branching fraction, $\mathcal{B}(B\to\etapr X_s)=(3.9\pm0.8\stat\pm0.5\syst\pm0.8\model)\times 10^{-4}$, for $2.0<p^*(\etapr)<2.7~\gevc$. We have also derived the $m(X_s)$ spectrum and found that the data tends to confirm models predicting a peak at high masses and seems to disfavor predictions based only on the diagram of Fig. \ref{fig:diagrams}(a,b) for which $m(X_s)$ peaks near 1.4-1.5~\gevcc \cite{ref:Datta}.\\
\indent Among the various theoretical conjectures to explain this
production, an $\etapr gg$
coupling due to the QCD anomaly has been widely suggested as a likely explanation. However, the $\etapr gg$ form factor initially proposed
\cite{ref:Atw} is disfavored by recent studies of the
inclusive production $\Upsilon(1S)\to\etapr X$
\cite{ref:KaganNew,ref:CLEOUps1}. A recently updated approach
\cite{ref:Fritzsch} exploiting the same \etapr gluon anomaly could
in principle account for the observed branching fraction and the $m(X_s)$
spectrum.

We are grateful for the excellent luminosity and machine conditions
provided by our \pep2\ colleagues.
The collaborating institutions wish to thank 
SLAC for its support and kind hospitality. 
This work is supported by
DOE
and NSF (USA),
NSERC (Canada),
IHEP (China),
CEA and
CNRS-IN2P3
(France),
BMBF
(Germany),
INFN (Italy),
NFR (Norway),
MIST (Russia), and
PPARC (United Kingdom). 
Individuals have received support from the Swiss NSF, 
A.~P.~Sloan Foundation, 
Research Corporation,
and Alexander von Humboldt Foundation.


\begin{thebibliography}{99}
\bibitem{ref:chargeconj}
Throughout this paper, whenever a mode is given, the charge conjugate state is also implied.
\end{thebibliography}
\end{document}